\documentclass[useAMS,usenatbib]{mn2e}
\pdfoutput=1
\usepackage{amssymb,amsmath}
\usepackage{epsfig,rotating,xspace,graphicx,subfigure}
\usepackage[3D]{movie15}
\usepackage{hyperref}
\newcommand{\msunyr}{\ensuremath{\mathit{M}_{\odot} \ {\rm yr}^{-1}}}   
\newcommand{\kms}{\ensuremath{{\rm km\,s^{-1}}}}                   
\newcommand{\msun}{\ensuremath{\mathit{M}_{\odot}}}   
\newcommand{\lsun}{\ensuremath{\mathit{L}_{\odot}}}                  

\newcommand{\lstar}{\ensuremath{\mathit{L}_{\star}}}                 
\newcommand{\mdot}{\ensuremath{\dot{M}}}                             
\newcommand{\mstar}{\ensuremath{\mathit{M}_{\star}}}                 
\newcommand{\teff}{\ensuremath{\mathit{T}_{\rm eff}}}                
\newcommand{\vinf}{\ensuremath{v_{\infty}}}                          

\newcommand{\K}{\ensuremath{\mathrm{K}}}                

\newcommand{\HI}{\ifmmode{{\rm H\,I}}\else{H{\sc$\,$i}}\fi\xspace}
\newcommand{\HII}{\ifmmode{{\rm H\,II}}\else{H{\sc$\,$ii}}\fi\xspace}
\newcommand{\HeI}{\ifmmode{{\rm He\,I}}\else{He{\sc$\,$i}}\fi\xspace}
\newcommand{\HeII}{\ifmmode{{\rm He\,II}}\else{He{\sc$\,$ii}}\fi\xspace}
\newcommand{\HeIII}{\ifmmode{{\rm He\,III}}\else{He{\sc$\,$iii}}\fi\xspace}
\newcommand{\FeIII}{\ifmmode{[{\rm Fe\,III}]}\else{[Fe{\sc$\,$iii}]}\fi\xspace}
\newcommand{\FeII}{\ifmmode{[{\rm Fe\,II}]}\else{[Fe{\sc$\,$ii}]}\fi\xspace}
\newcommand{\NiII}{\ifmmode{[{\rm Ni\,II}]}\else{[Ni{\sc$\,$ii}]}\fi\xspace}

\title{The three-dimensional structure of the Eta~Carinae Homunculus \thanks{\noindent Based on observations made with ESO Telescopes at the La Silla Paranal Observatory under programme ID 088.D-0873(A) (PI Groh).}}
\author[W. Steffen et al.]{W. Steffen$^{1}$\thanks{e-mail: \texttt{wsteffen@astro.unam.mx}}, M. Teodoro$^{2}$\thanks{CNPq/Science without Borders Fellow}, T.~I. Madura$^{2}$\thanks{NASA Postdoctoral Program Fellow}, J.~H. Groh$^{3}$, T.~R. Gull$^{2}$,
\newauthor A. Mehner$^{4}$, M.~F. Corcoran$^{5,6}$, A. Damineli$^{7}$, K. Hamaguchi$^{5,8}$\\\\
$^{1}$Instituto de Astronom\'{\i}a, Universidad Nacional Aut\'onoma de
M\'exico, Apdo Postal 106, Ensenada 22800, Baja California, M\'exico\\
$^{2}$Astrophysics Science Division, Code 667, NASA Goddard Space Flight Center, Greenbelt, MD 20771, USA \\
$^{3}$Geneva Observatory, Geneva University, Chemin des Maillettes 51, CH-1290, Sauverny, Switzerland \\
$^{4}$ESO, Alonso de Cordova 3107, Vitacura, Santiago de Chile, Chile \\
$^{5}$CRESST and X-ray Astrophysics Laboratory, NASA Goddard Space Flight Center, Greenbelt, MD 20771, USA \\
$^{6}$Universities Space Research Association, 7178 Columbia Gateway Drive, Columbia, MD 21046, USA \\
$^{7}$Instituto de Astronomia, Geof\'{\i}sica e Ci\^{e}ncias Atmosf\'{e}ricas, Universidade de S\~{a}o Paulo, Rua do Mat\~{a}o 1226, Cidade Universit\'{a}ria, \\ S\~{a}o Paulo, 05508-900, Brazil \\
$^{8}$Department of Physics, University of Maryland, Baltimore County, 1000 Hilltop Circle, Baltimore, MD 21250, USA}

\begin{document}

\date{Accepted 2014 May 29. Received 2014 May 29; in original form 2014 April 3}

\pagerange{\pageref{firstpage}--\pageref{lastpage}}

\maketitle
\label{firstpage}

\begin{abstract}
\noindent
We investigate, using the modeling code {\sc SHAPE}, the  three-dimensional structure of the bipolar Homunculus nebula surrounding Eta Carinae as mapped by new ESO VLT/X-Shooter observations of the H$_\mathrm{2}$ $\lambda=2.12125\mu$m emission line. Our results reveal for the first time important deviations from the axisymmetric bipolar morphology: 1) circumpolar trenches in each lobe positioned point-symmetrically from the center and 2) off-planar protrusions in the equatorial region from each lobe at longitudinal ($\sim55^\circ$) and latitudinal ($10^\circ-20^\circ$) distances from the projected apastron direction of the binary orbit. The angular distance between the protrusions ($\sim110^{\circ}$) is similar to the angular extent of each polar trench ($\sim130^{\circ}$) and nearly equal to the opening angle of the wind-wind collision cavity ($\sim110^{\circ}$). As in previous studies, we confirm a hole near the centre of each polar lobe and no detectable near-IR H$_\mathrm{2}$ emission from the thin optical skirt seen prominently in visible imagery. We conclude that the interaction between the outflows and/or radiation from the central binary stars and their orientation in space has had, and possibly still has, a strong influence on the Homunculus. This implies that prevailing theoretical models of the Homunculus are incomplete as most assume a single star origin that produces an axisymmetric nebula. We discuss how the newly found features might be related to the Homunculus ejection, the central binary and the interacting stellar winds.
\end{abstract}


\section{Introduction}
\label{intro}

Massive stars have major impacts on their host galaxies via the input of ionizing photons, energy, and momentum into the interstellar medium. They also contribute to the chemical evolution of the Universe via the ejection of enriched elements through stellar winds and supernovae. Our current understanding is that some massive stars undergo violent ejections of a fraction to several-tens of solar masses as they evolve through the luminous blue variable (LBV) stage \citep{langer94,hd94,so06}. The mechanisms responsible for these eruptive events, as well as their frequency and the total amount of mass lost, are not currently understood.

Eta~Carinae is the closest and most studied example of an object that underwent a `Great Eruption' \citep{davidson12}. At least 10\msun, and potentially more than 40\msun, were ejected during this event \citep{morris99,smith03a,smith06,gomez10}, causing the formation of a dusty bipolar nebula known as the Homunculus \citep{gaviola50}. Analysis of the proper motion of the Homunculus shows that the eruption happened in the 1840s \citep{currie96,morse01}. This, combined with its relative proximity (2.3~kpc, \citealt{allen93,smith06}), means that the Homunculus's structure can be investigated in an unparalleled level of detail. This provides valuable insights into the mechanisms responsible for LBV eruptions and their role in the evolution of the most massive stars.

Eta~Carinae itself is a massive, highly eccentric ($e \sim 0.9$) binary system \citep{dcl97} with a combined luminosity and mass of $\lstar\simeq5\times10^6~\lsun$ \citep{cox95,davidson97} and $\mstar> 110~\msun$ \citep{hillier01}, respectively. The LBV primary is responsible for most of the system luminosity (and thus mass), with a mass-loss rate of $8.5\times10^{-4} \msunyr$ and wind terminal velocity of $\sim420~\kms$ (\citealt{ghm12}; see also \citealt{hillier01,hillier06}). This extremely dense wind causes the photosphere to be formed in the wind, and one infers an effective temperature of $\teff\simeq9400~\K$ at optical depth $\tau = 2/3$ \citep{hillier01,hillier06,ghm12}. Only indirect constraints are available for the companion star, which likely has a temperature of $\teff\simeq36,000-41,000~\K$, luminosity $10^5 \lsun \la \lstar \la 10^6 \lsun$ \citep{mehner10}, wind terminal speed $\vinf\sim3000~\kms$, and mass-loss rate $\mdot \sim1.4 \times 10^{-5}~\msunyr$ \citep{pc02,parkin09}.  Currently, the companion has a significant impact on the wind of the primary \citep{pc02,okazaki08,parkin09,mg12,gmo10,gnd10,ghm12,gmh12,madura12,madura13}. However, it is unclear whether the companion had any role in triggering the eruption or shaping the Homunculus.

Numerous theoretical studies have investigated the formation of the Homunculus. These studies fall mainly into six categories: interacting winds \citep{frank95,frank98,dwarkadas98,langer99,gonzalez04a,gonzalez04b}, binary interaction \citep{soker01,soker04,soker07,kashi10}, a radiatively-driven wind from a rapidly-rotating star \citep{owocki97,maeder01,dwarkadas02,owocki03,owocki05}, a thermally-driven magnetohydrodynamic (MHD) rotating wind \citep{matt04}, explosive mass loss from a rapidly-rotating star (e.g. \citealt{smithtownsend07}), and stellar mergers \citep{gallagher89,iben99,morris06}. The interacting winds scenario requires a pre-existing slow and dense equatorial torus to pinch the waist of the bipolar structure. However, observations show that the thin equatorial disc or `skirt' seen in optical images of the Homunculus is no older than the bipolar lobes and that there is not enough mass in the skirt to shape the overall morphology of the nebula \citep{SG98,davidson01,morse01,smith06}. The MHD model of \citet{matt04} requires a huge field strength ($\gtrsim 2.5 \times 10^{4}$~G at the equator) to shape the large amount of mass ejected, and it is unclear if such a field was available. Models employing radiatively-driven winds from a rapidly-rotating star have difficulties reproducing simultaneously the bipolar structure and the thin equatorial skirt. The large amount of mass expelled over such a short period of time also cannot be readily explained by the standard line-driven-wind formalism, and may even surpass the capability of a super-Eddington continuum-driven wind \citep{owocki04, smithtownsend07}. However, a continuum-driven \emph{explosion} from a rapidly-rotating single star may explain both the bipolar shape and thin equatorial skirt \citep{smithtownsend07}. Stellar mergers have difficulty accounting for previous and subsequent large eruptions, such as the smaller eruption in 1890 that formed the Little Homunculus \citep{ishibashi03,smith04}. While mergers may seem unlikely, they also cannot yet be fully ruled out.

Besides the single-star axisymmetric model of \citet{smithtownsend07}, another possibility for explaining the formation and shape of the Homunculus involves some sort of binary interaction, perhaps during periastron passage \citep{kashi10,smith11}. Possible evidence for such a scenario is the correlation of periastron passage with the peaks observed in Eta~Carinae's light curve in the decades leading up to the Great Eruption (see fig.~2 of \citealt{damineli96} and also \citealt{smithfrew11}). Moreover, it was recently found that the orbital axis (orthogonal to the orbital plane and through the system centre of mass) of the Eta~Carinae binary is closely aligned in three-dimensional (3D) space with the inferred polar symmetry axis of the Homunculus \citep{madura12}. Unfortunately, all current models for the Homunculus that involve binary interaction are phenomenological and rely on assumptions such as strong accretion by the secondary star. These simple models are incapable of predicting the exact shape and degree of asymmetry in any resulting nebula \citep{soker01}. Numerical-hydrodynamical simulations are required.

If the binary companion had any influence on the formation or shaping of the Homunculus, it should have left an imprint in the Homunculus's morphology. There are at least two ways the binary companion and Homunculus's formation might be related, and both could be relevant. First, the companion may be directly responsible for triggering the eruption itself, say via a stellar collision \citep{smith11}. Second, the companion could influence the shape of the nebula during and/or after any explosion, via a collimated fast wind like that proposed by \citet{soker01}, or simply via its extremely fast, low-density wind and orbital motion.

Of course, any theoretical model for the Homunculus can only be as good as the observations on which it is based. Several observational studies, using mainly long-slit optical spectroscopy and mid-infrared imaging, have investigated the 3D structure of the Homunculus. Ground-based optical spectroscopy at medium spectral resolution has long been employed to measure doppler-shifts of emission lines formed in the Homunculus lobes, showing that the nebula is bipolar with an approaching  south-east (SE) lobe, a receding north-west (NW) lobe, and a thin equatorial skirt \citep{thackeray51,thackeray56a,thackeray56b,thackeray61,meaburn87,meaburn93,hillier92,allen91,allen93,hillier97,currie96,meaburn99}. The improved spatial resolution of {\it Hubble Space Telescope} ({\it HST})/Space Telescope Imaging Spectrograph (STIS) optical observations allowed further detailed studies of the shape and orientation of the Homunculus lobes \citep{davidson01}. The study of near-IR (NIR) emission lines discovered the presence of molecular gas (H$_\mathrm{2}$; \citealt{smithdavidson01}), making it possible to trace the structure of the optically-obscured back-side of the Homunculus \citep{smith02,smith06,T08}. \citet{smith03a} proposed that the H$_\mathrm{2}$ gas is likely associated with the outer, cool dust shell seen in mid-IR images \citep{hackwell86,smith95,SGK98,polomski99,chesneau05}, which traces the majority ($\sim 90$\%) of the Homunculus mass (15--45~\msun; \citealt{smith03a,smith06,gomez10}). NIR observations \citep{smith02} also showed that the H$_\mathrm{2}$~$v=1-0$~S(1) emission originates in a shell located exterior to the region of \FeII and \NiII emission investigated by \citet{davidson01}.

Subsequent ground-based, high-spectral-resolution observations of H$_\mathrm{2}$ and \FeII lines provided the most detailed view of the Homunculus to date \citep{smith06}. This study confirmed the double-shell structure of the Homunculus, with inner \FeII and outer H$_\mathrm{2}$ shells. It also showed that about 75\% of the Homunculus mass is located at high latitudes (above 45\degr). In addition, more than 90\% of the kinetic energy was released at these high latitudes during the Great Eruption, which seems to have had a short duration ($\la 5$~years) given the thinness of the H$_\mathrm{2}$ shell \citep{smith06}. However, these observations had limited spatial coverage, using only five long-slits aligned along the major polar axis of the Homunculus. This provides insufficient information for a complete 3D study of the structure of the Homunculus. Moreover, all previous observational studies of the Homunculus focused on its large-scale axisymmetric shape in the context of a single-star model. Departures from axisymmetry and the role of the companion are either only briefly mentioned or neglected. Detailed knowledge of the 3D structure of the Homunculus, including small-scale departures from axisymmetry, is crucial for any proper theoretical modeling of its formation and the Great Eruption.

Using an extensive new set of ESO Very Large Telescope (VLT)/X-shooter spectral mapping observations of the entire Homunculus, we investigate the Homunculus's full 3D structure, providing a more detailed view of its shape. In particular, we probe deviations from axisymmetry to determine whether the companion star might have played a role in shaping the Homunculus. We analyze the X-shooter data cubes using the {\sc SHAPE} software \citep{SKWMM11} and generate, for the first time, a complete 3D model of the Homunculus that includes smaller-scale features. In this initial investigation we focus on the H$_2$~$v=1-0$~S(1) emission line, which traces the fronts and backs of the Homunculus's polar lobes \citep{smith06}.

This paper is organized as follows. In Sect.~\ref{observations.sec} we present our new observational mapping data obtained with VLT/X-shooter. Sect.~\ref{model.sec} describes the {\sc SHAPE} modeling procedure and Sect.~\ref{results.sec} the results. We discuss the results in Sect.~\ref{disc.sec} before summarizing our conclusions in Sect.~\ref{conclusions.sec}.

\begin{figure}
  \begin{center}
  \includegraphics[width=0.4\textwidth]{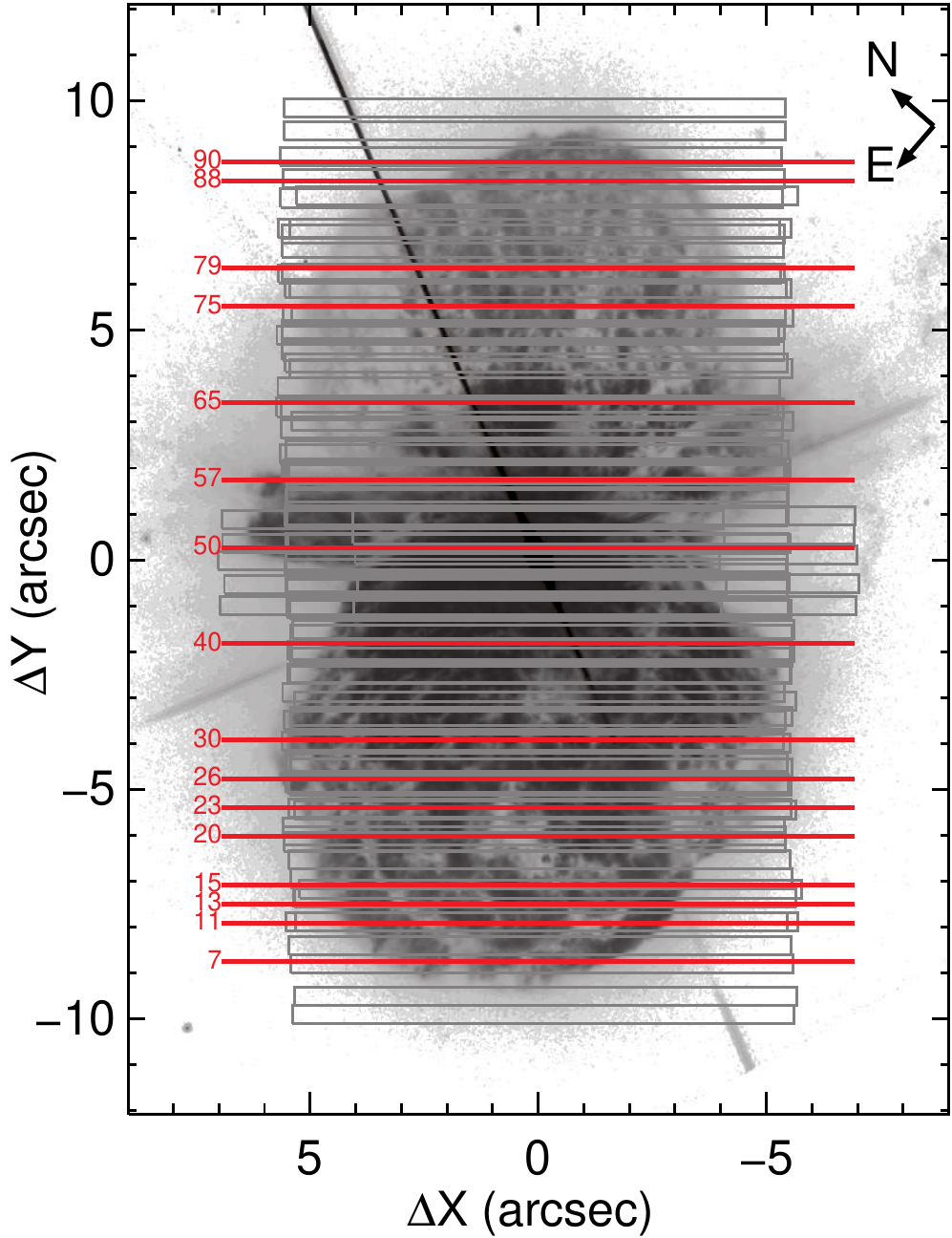}
  \caption{Mapping position of each slit superimposed over an \textit{HST}/ACS F550M image of the Homunculus obtained on 4 August 2006 (HST Proposal ID 10844, PI: K.~Davidson; \citealt{martin06}). The original image has been rotated by 49$^\circ$ counter-clockwise and corrected for the mean expansion of the nebula between 2006.6 and 2012.2 $\approx0.20\arcsec$ (Smith \& Gehrz 1998). Horizontal red lines mark the numbered slit locations where select position-velocity (P-V) images used in this paper were extracted from the observational data cube (see Fig.~\ref{model_pv1.fig}).}
  \label{fig:refimage}
  \end{center}
\end{figure}

\section{Observations, data reduction and data cube assembling}
\label{observations.sec}

A full spectral mapping (2990~\AA\ through 24790~\AA) of the entire Homunculus nebula ($14\arcsec\times20\arcsec$) was obtained on the nights of 9 and 10 March 2012, using the ESO VLT/X-shooter spectrograph (ESO program ID 088.D-0873(A)). A detailed description of the instrument's capabilities and configurations is given in \citet{vernet11}. Since we focus on the H$_2$~$v=1-0$~S(1) $\lambda = 2.12125$~$\mu$m\footnote{All wavelengths in this paper refer to measurements in air.} emission line to model the Homunculus, we discuss here only the parameters, processing, and reductions for the X-shooter NIR arm.

The data were obtained using the $0.4\arcsec\times11\arcsec$ slit, which resulted in a spectrum with $R=\lambda/\delta\lambda=11300$ and a sampling of 2 pixels per FWHM. The mapping of the nebula was done with the slit oriented at position angle P.A.=$-41^\circ$, i.e. perpendicular to the projected major axis of the Homunculus. A total of 92 dithered positions along the nebula was used in the full mapping (Fig.~\ref{fig:refimage}). The spatial coverage was from $-6.93\arcsec$ to $+6.93\arcsec$ from the central source in the X direction, and from $-10.02\arcsec$ to $+9.93\arcsec$ in the Y direction. The individual exposure times ranged from 0.06651~seconds close to the central source to 30~seconds in the lobes in order to not overexpose. Multiple images at each position were obtained and combined. The total exposure time thus ranged from 30~seconds close to the star to 150~seconds in the lobes.

Data processing and reduction were performed using the X-shooter pipeline (cpl-6.1.1 and xsh/1.5.0; \citealt{2006SPIE.6269E..80G}) for which the standard processing includes removal of dark current, order identification and tracing, correction for pixel-to-pixel sensitivity variations, dispersion solution, master response, and merging of all orders into a single wavelength calibrated 2D spectrum (for each slit position). In the spectral direction, each reduced 2D spectrum ranges from 9940~\AA\ to 24790~\AA~with a dispersion of 1~\AA~per pixel, whereas in the spatial direction, the final scale for the NIR data is $0.21\arcsec$ per pixel.

The assembling of the final data cube was achieved by creating a grid with spatial dimensions set by the maximum displacement between the offset positions along slit width and length. Each spatial pixel in the data cube grid was set to a size of $0.21\arcsec\times0.21\arcsec$. The size of each pixel in the spectral direction was not changed. Therefore, in the final data cube, each voxel has $0.21\arcsec\times0.21\arcsec\times1.0$~\AA. In the case where slits overlap each other, we adopted the average value for each overlapping pixel.

\begin{figure*}
  \begin{center}
  \includegraphics[width=0.7\textwidth]{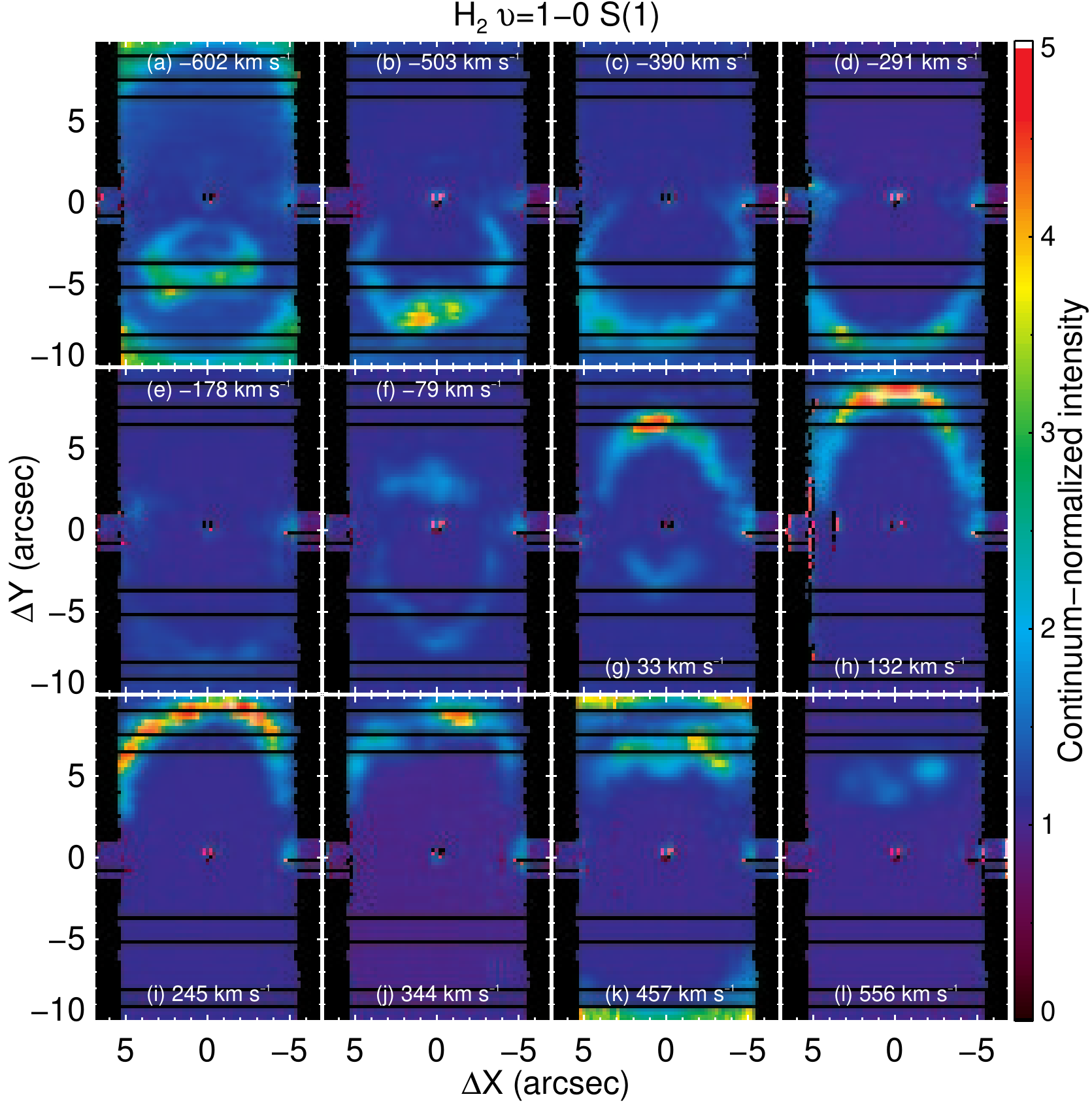}
  \caption{Iso-velocity images of H$_{2}$~$v=1-0$~S(1) at $2.12125$~$\mu$m. The structures seen in emission in each slice were mapped both in the spatial and spectral directions, allowing the reconstruction of the 3D shape of the emitting volume. The black horizontal lines across the images are regions where the spacing between adjacent slits was larger than the slit width and, therefore, contain no data.}
  \label{fig:H2}
  \end{center}
\end{figure*}

As an illustration of the high quality of the final assembled data cube, Fig.~\ref{fig:H2} shows H$_2$ iso-velocity maps for which the spectra were normalized by the adjacent continuum to highlight the structures in emission (corresponding to material in the Homunculus lobes). Although line emission from the backs of the lobes is weaker than that coming from the fronts (see Figs.~\ref{fig:H2}\textit{e}, \textit{f}, and \textit{l}), we were still able to map them.

\section{3D structure model}
\label{model.sec}

With the 3D morpho-kinematic modeling we wish to chiefly address two questions: 1) Does the small-scale structure of the Homunculus, beyond the cylindrical-symmetric component of earlier models, include information on the details of the ejection mechanism? 2) Are there symmetries that might indicate a relation to the central binary interaction?

\begin{figure*}
\centering \includegraphics[width=150mm]{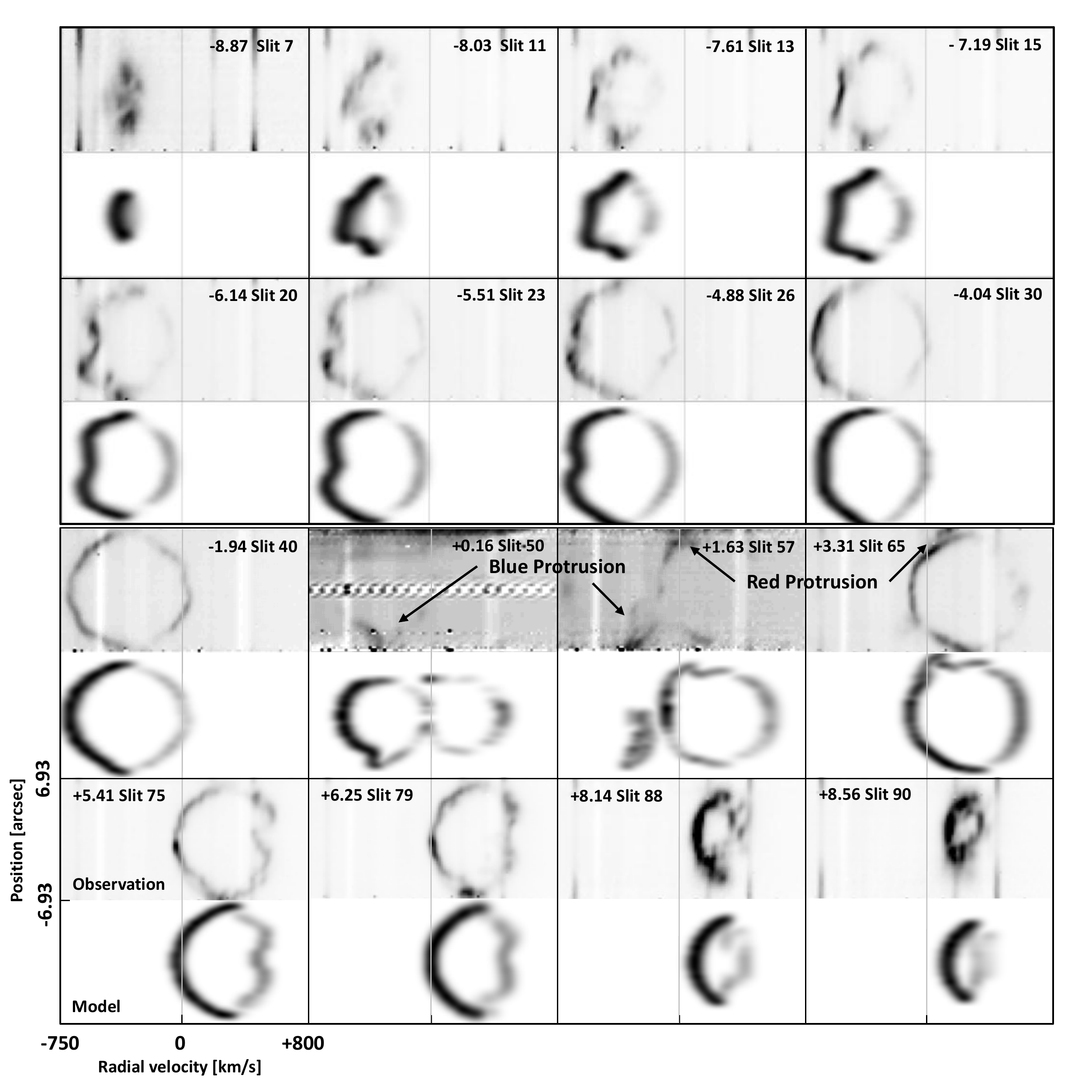}
\caption{Observed and model P-V images. The observational P-V images (upper panel in each P-V image pair), selected from the approximately one hundred images available, are most representative of the spatial-velocity changes as the slit moves across the Homunculus along its main axis, with the slit direction perpendicular to the polar axis of the Homunculus. The offset position of the numbered slit from the central source, in arcseconds, is indicated for each pair of P-V images (see Fig.~\ref{fig:refimage}). The spatial and spectral ranges of the P-V diagrams are $-6.93\arcsec$ to $+6.93\arcsec$ and $-750$~km s$^{-1}$ to $+800$~km s$^{-1}$, respectively.}
\label{model_pv1.fig}
\end{figure*}

Previously constructed morpho-kinematic models of the Homunculus assumed axisymmetry, yielding only its overall shape (e.g. \citealt{davidson01,smith06,T08}), although departures from true axial symmetry are known \citep{morse98,smith06}. In this paper we consider three refinements to previous models. The first is the discrepancy between models based on Doppler-velocity measurements and those based on images. Second, we attempt to reconstruct some of the small-scale structures in the polar regions. Third, we include two newly-found kinematic features near the equatorial region of the Homunculus. As in \citet{smith06}, we find no detectable NIR H$_{2}$ emission from the thin equatorial skirt that is prominent in optical images of the Homunculus. We therefore model only the main bipolar structure.

We first consider the overall structure of the Homunculus. Some papers present image- and velocity-based models in the same work (e.g. \citealt{davidson01}). The systematic discrepancy between models based on images and spectroscopy is that the spectroscopic reconstructions from molecular hydrogen yield a slimmer Homunculus than the image-based models. Several possible explanations come to mind for such a method-dependent difference in the reconstructions. First, the differences could be intrinsic, since the image-based reconstructions are based on continuum scattered by dust, whereas the Doppler-reconstructions use line emission that might originate in spatially distinct regions. Furthermore, the fact that Eta~Carinae is a binary with a highly elliptical orbit and interacting winds could produce a symmetry that is non-cylindrical.

The impression from the optical images, defined largely by dust-scattered continuum light from Eta~Carinae interior to the Homunculus shell, is that a relatively thin, clumpy surface defines the location of the scattering dust. One is left with the impression that the bulk of the H$_{2}$ emission is just outside this surface. Molecular hydrogen is thus expected to be located mainly {\em outside} the dust shell and hence also outside the continuum emission. However, reconstructions from molecular line emission show a slimmer Homunculus than expected and hint toward a problem in the assumptions of the velocity field.

Reconstructions based on Doppler-velocity measurements of expanding nebulae often assume homologous expansion, i.e. radial expansion with the magnitude of the velocity increasing linearly with distance. This is a very stringent assumption and can be met only under certain conditions, and only to some level of accuracy. Hydrodynamical simulations of wind-blown bipolar nebulae show that substantial distortion can occur in 3D reconstructions if deviations from a homologous expansion are not taken into account \citep{SKG09}. We therefore explore whether hydrodynamically-plausible deviations from a homologous expansion can lead to a more consistent 3D reconstruction of the large-scale structure.

In addition to the large-scale structure, we attempt to reconstruct the most salient small-scale features based on the molecular hydrogen Doppler-velocity measurements presented in Figs~\ref{fig:H2} and \ref{model_pv1.fig}. Here it is important to note that we reconstruct a single continuous surface of each lobe and we assume that the deviations from a homologous expansion are only large-scale, such that the local structures are reasonably well mapped by the velocity measurements.

As will be discussed later, some smaller-scale structures in the polar regions of the Homunculus may be caused by hydrodynamical instabilities, that at some time in the past must have introduced local velocity disturbances. If instabilities have been active in the relatively distant past and are now only passive, then the local deviations from homologous expansion will be small. Detailed numerical hydrodynamic simulations are needed to test the accuracy of the assumption of locally homologous expansion in the complex polar regions of the Homunculus.

Assuming that deviations from a homologous expansion are only large scale, we first investigate the large-scale Homunculus shape and then the small-scale structure of the polar and equatorial regions. Afterward, we investigate the effects of dust in order to understand better the observed brightness distribution in the H$_2$ data cube and its representation as position-velocity (P-V) diagrams.

\subsection{Modeling method}
\label{method.sec}

\begin{figure*}
\centering
\subfigure{
\includegraphics[width=80mm]{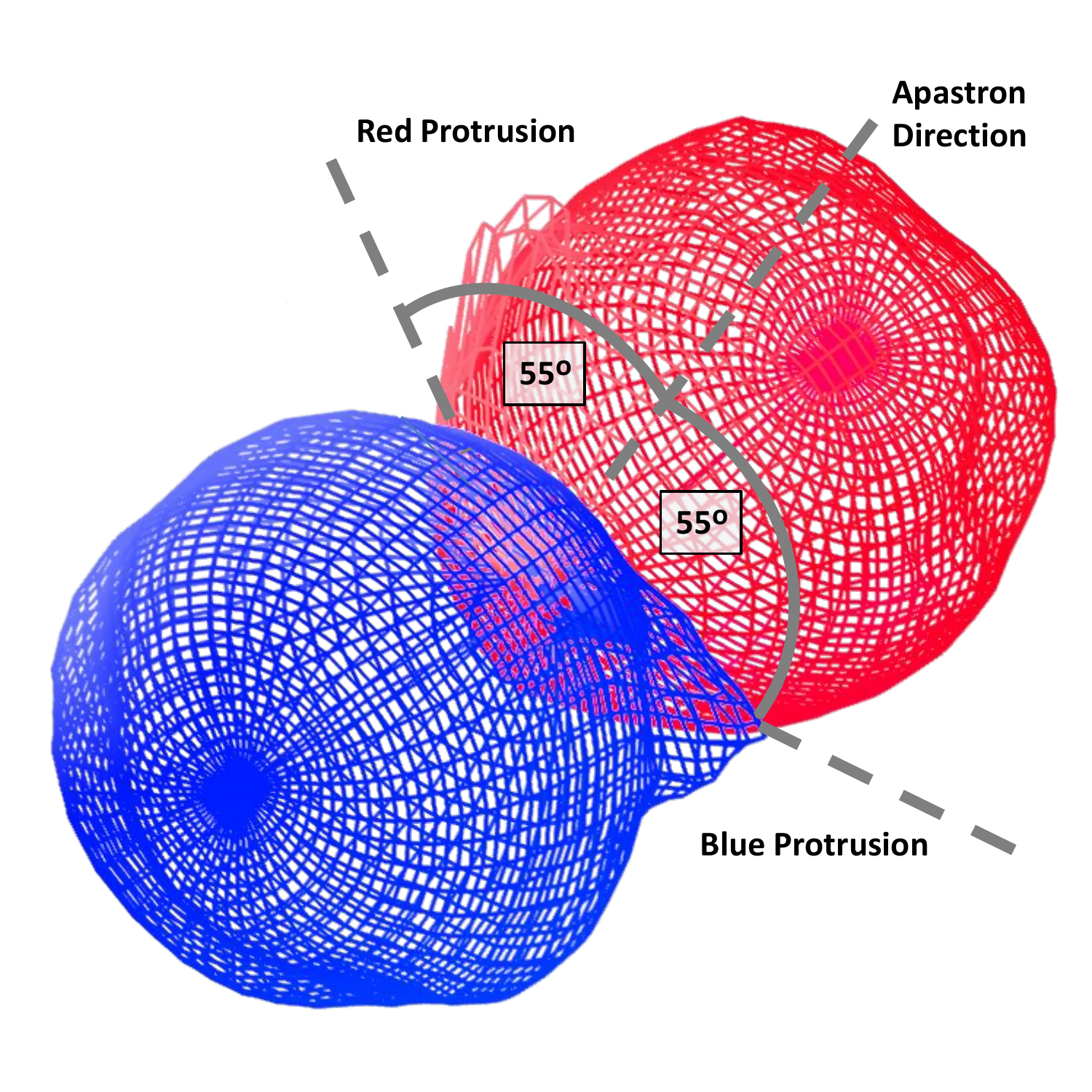}
\label{fig4a}}
\subfigure{
\includegraphics[width=80mm]{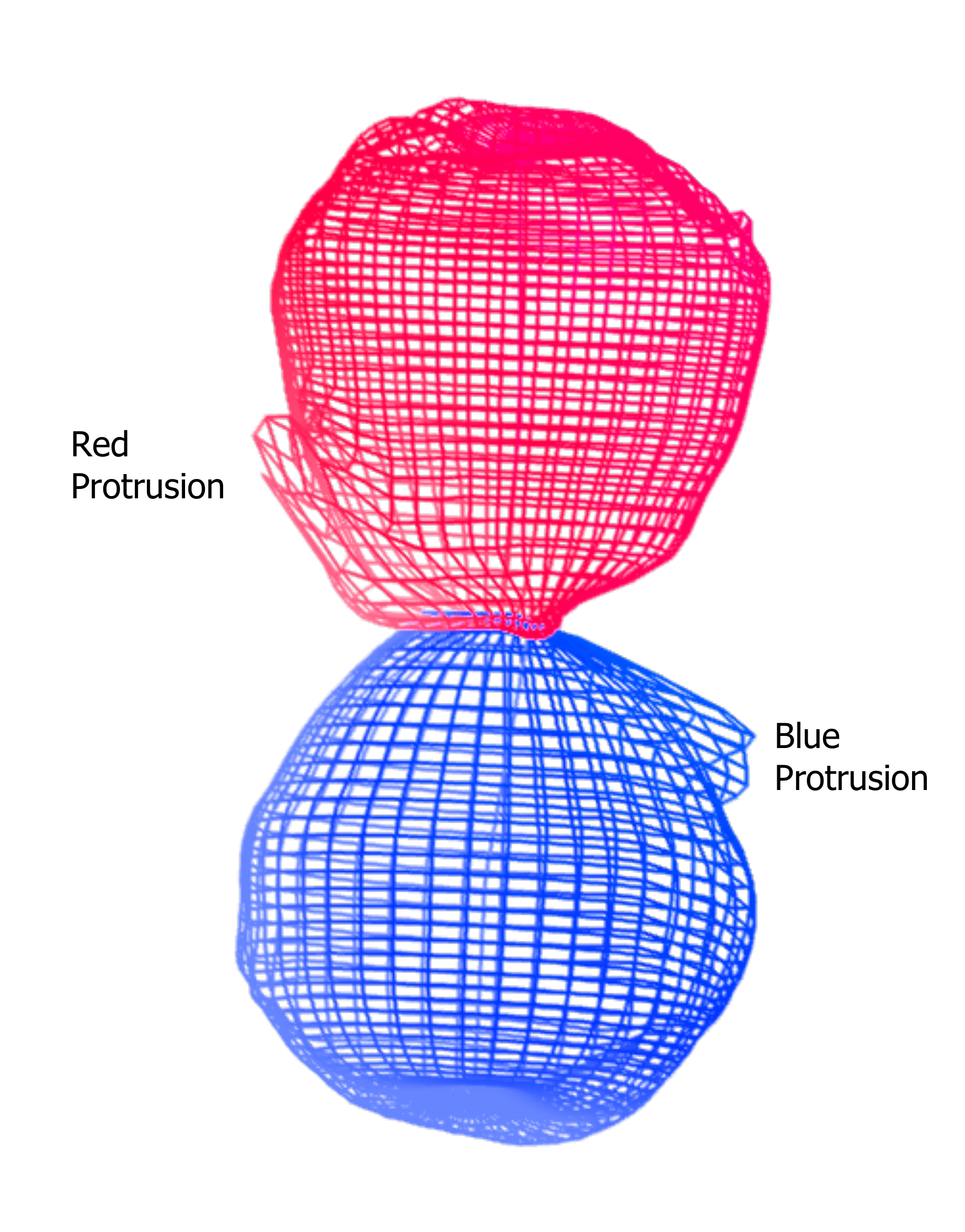}
\label{fig4b}}
\caption{The 3D mesh structure of the {\sc SHAPE} Homunculus model. For clarity, the mesh resolution has been reduced. \emph{Left}: Orientation as viewed from Earth, with North up and East to the left. The directions, as projected onto the equatorial plane, of orbital apastron and the new near-equatorial kinematic features (e.g. the Red and Blue Protrusions) are marked with grey lines. \emph{Right}: Similar to left, but for a non-inclined, non-rotated view where the polar axis is parallel to the plane of the page, showing that the Red and Blue Protrusions extend from the NW-red and SE-blue lobes, respectively, and do not lie in the equatorial plane.}
\label{dents.fig}
\end{figure*}

We constructed the 3D model using the morpho-kinematic and radiative transfer code {\sc SHAPE} \citep{SKWMM11}. Based on images and spectroscopic data, {\sc SHAPE} has been used extensively for modeling the complex 3D structure of planetary nebulae (e.g. \citealt{GD12,clark13}), novae \citep{R13}, radiation transfer in dusty pre-planetary nebulae \citep{KKS13}, quark-nova ejecta \citep{ouyed12}, and the complex planetary nebula Hubble~5 \citep{LGSRR12}, which in shape is rather similar to the Homunculus. The model is constructed interactively with 3D mesh structures (see Fig.~\ref{dents.fig}). A velocity field and a brightness distribution is then assigned. We use the physical renderer of {\sc SHAPE} which allows radiative transfer with dust opacity. Radiative transfer is helpful for modeling of the small-scale structures in the Homunculus since the effects of continuum absorption by dust cannot be neglected. Therefore, we introduce a dust shell that is basically a copy of the H$_2$ emission shell, but with a slightly smaller size so that they only partially overlap.

For this work the details of the dust properties are not relevant, but for completeness we mention that we use optical properties appropriate for silicate dust with a power-law size distribution between 0.5 and 2.5 microns, a power-law index of 3.5, and optical properties according to \citet{K08}. The H$_2$ emissivity and dust density are assumed to be a smooth distribution that decreases slowly with distance. Therefore, the resultant, often strong, projected image model brightness variations are mainly due to the integrated effective path-length along the line of sight, rather than intrinsic variations.

For the spectral line computation we use a gaussian line profile at a wavelength of 2.1218~$\mu$m with a $\sigma = 10^{-4}$$\mu$m, i.e. a FWHM of $33$~km s$^{-1}$, which ensures the profile is properly sampled with 128 channels over a full spectral range of $\pm 700$~km s$^{-1}$ around the local standard of rest.

Small-scale structure is added using various standard tools that modify the shape of the initially spherical 3D mesh. The general modeling procedure is described on the support website of {\sc SHAPE} ({\em www.astrosen.unam.mx/shape}). Modeling the location of individual small-scale features is based on the following assumption: the Homunculus can be approximated as a surface structure and that the image together with the expected velocity field yields a one-to-one mapping between the projected observed position and velocity to the position on the Homunculus surface (e.g. \citealt{S04}; \citealt{M05}, \citealt{SKG09}).

The model P-V diagrams (Fig.~\ref{model_pv1.fig}) and images have been individually normalized and use a linear grayscale coding. As this paper focuses on the structure of the Homunculus, the overall relative brightness distribution is used only as a general guideline.

\section{Results}
\label{results.sec}

In this section we describe the results of modeling the small-scale structures of the bipolar Homunculus using the H$_2$ observations presented in this paper, as well as some basic results regarding the location of the scattering dust relative to the emitting H$_2$. For simplicity, we use the terms `blue lobe' and `red lobe' when referring to the Homunculus's approaching (SE) and receding (NW) lobes, respectively.

\subsection{Velocity field}
\label{velocity.sec}

As mentioned, the assumption of a homologous expansion for the Homunculus leads to a reconstruction of the structure along the line of sight that is inconsistent with the continuum image and the expectation that the molecular line emission originates outside of the dust. We find that within the precision of the data and the reconstruction method, a non-homologous expansion is only required for the blue lobe. For the blue lobe, the model includes a non-homologous velocity field in the form of a small poloidal velocity component with a latitudinal variation \citep{SKG09}. Here the velocity field is divided into two locally perpendicular components, i.e. a radial and poloidal velocity. The radial velocity $v_{\rm r} (r) = 680~{\rm km \ s}^{-1}~(r/3.4\times 10^{15}~{\rm m})^{1.2}$ is a function that is slightly non-linear, increasing faster than linearly with distance $r$ from the central star. The poloidal component $v_{\phi} (\phi)$ has a broad peak at mid angles ($\approx$ 45$^\circ$) between the equator and pole of the blue lobe, going slowly to zero at the equator and pole. The peak poloidal velocity amounts to $\sim 50$~km~s$^{-1}$ and is hence only of order 10\% of the peak expansion velocity of the Homunculus. For the red lobe the radial velocity is $v_{\rm r} (r) = 600~{\rm km \ s}^{-1}~(r/3.4\times 10^{15}~{\rm m})$ and hence linear, but slightly smaller than that of the blue lobe. No poloidal component was necessary for the red lobe. We find that these small corrections solve the problem of different reconstructions based on images or Doppler-measurements, yielding a large-scale cylindrical outline that is consistent with both types of observations.

The precision of the exponent $1.2$ in the radial component and a peak $\sim 50$~km~s$^{-1}$ for the poloidal components of the velocity in the blue lobe is uncertain. These values should be taken only as an clear indication that there is some large-scale deviation from a homologous expansion, if the systematic deviation between Doppler-velocity and image based reconstructions is not real.

In the kinematic analysis of hydrodynamic simulations of wind-driven bipolar nebulae presented by \citet{SKG09}, the poloidal component points away from the pole, while here we find that the direction is toward the pole, implying a recollimation. While the effect is only a small percentage of the total velocity, it is large enough to change the overall shape of the reconstructed Homunculus. While the study by \citet{SKG09} was aimed at planetary nebulae driven by a continuous stellar wind, in Eta~Carinae an explosive expansion and different environmental conditions might produce the inverted poloidal velocity component. Such a flow is obtained for instance in a Cant\'o-Flow, where a wind is shocked on the inside of a lobe in an external density gradient and is redirected toward the pole of the lobe \citep{CR80}. New hydrodynamical simulations might shed light on the mechanisms responsible for this deviation from homologous expansion, but are outside the scope of this paper and must be left for future research.

\begin{table*}
\begin{center}
\begin{tabular}{lrrrr}
\hline
Feature & Longitude & $\Delta$ Longitude & Latitude & $\Delta$ Latitude  \\
\hline \hline
Blue trench   & 180   & 130   &  75   & 10 \\
Red trench    & 0     & 130   &  75   & 10 \\
Blue ``hole"     & 0     & N/A   &   8   & 15 \\
Red ``hole"      & 180   & N/A   &   8   & 12 \\
Blue Protrusion              & -55   & 30    &  10   &  7 \\
Red Protrusion               & 55   & 30    &  10-30& 20 \\
\hline
\end{tabular}
\caption{Summary of the approximate geometric properties of the small-scale features identified in the 3D model. Angles are measured in degrees. Longitude is with respect to the apastron direction of the binary star and positive is counter-clockwise from Earth's point of view. Here $\Delta$ refers to the angular spatial extent of a particular feature, not the uncertainty in the measurement. The uncertainties in the angular values are estimated to be of order $10-20$\%. In the case of the latitude of the polar holes, the latitudinal angle is given as the deviation from the main axis of the lobes in the model. The uncertainty of this value is $\approx 2^{\circ}$.}
\label{model.tab}
\end{center}
\end{table*}

\subsection{Small-scale structures}
\label{smallscale.sec}

The Homunculus is highly structured on small scales at visible and IR wavelengths, and departures from axisymmetry, especially in the blue lobe, have been known for some time \citep{morse98,smith06}. Filamentary dust structures highlight the observed speckled structure of the lobes (Fig.~\ref{fig:refimage}). Furthermore, there are apparently radially-expanding flat spikes in the equatorial plane and polar ``holes" in the lobes \citep{SGK98}.


\begin{figure}
\centering \includemovie[
     3Dviews2=views.tex,
        toolbar, 
        label=Fig5_animation.u3d,
     text={\includegraphics[width=80mm]{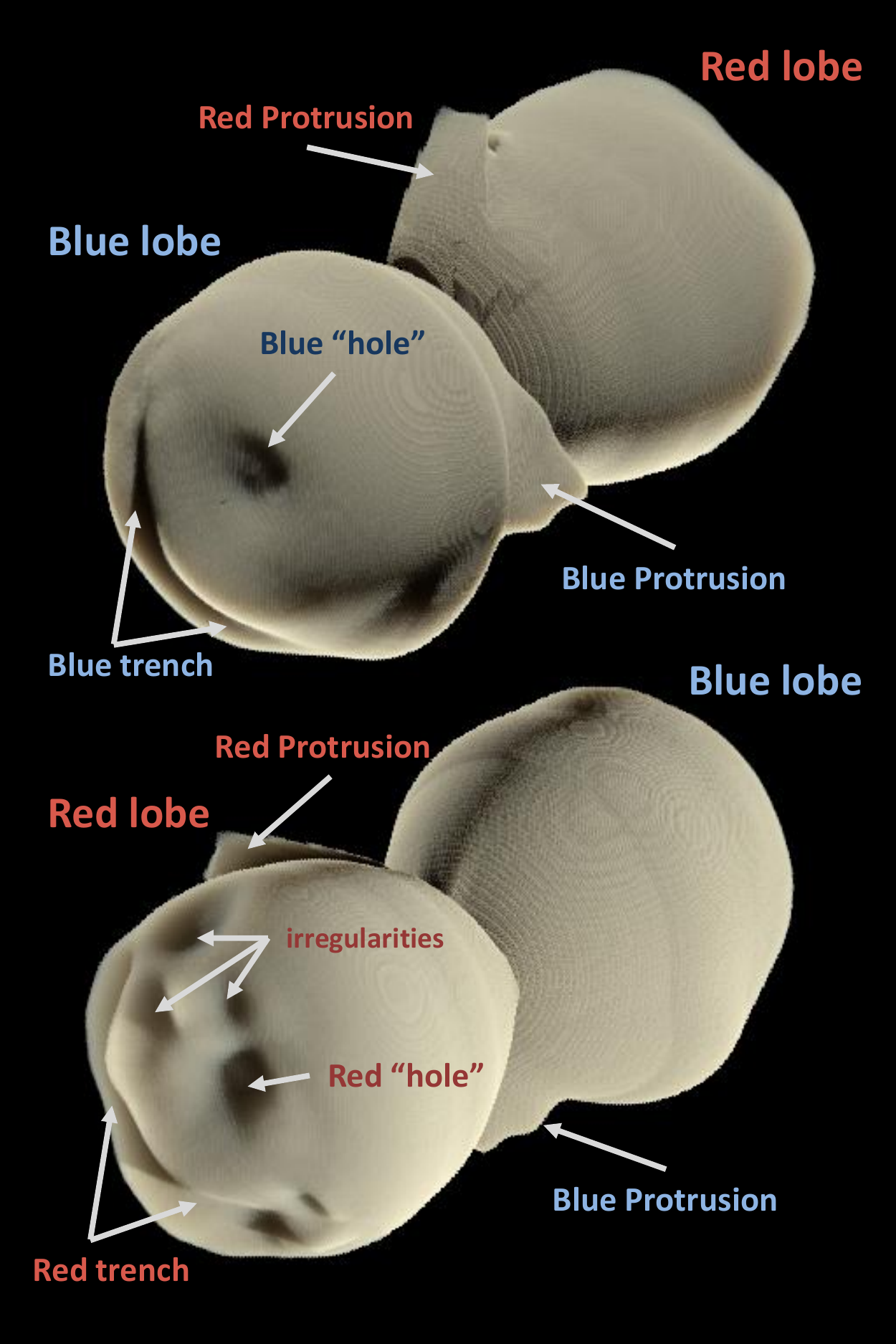}},
        3Daac=60.000000, 3Droll=127.43660402458738, 3Dc2c=-0.06051350384950638 0.6740224957466125 0.7362279891967773, 3Droo=9.31, 3Dcoo=-0.06521779298782349 0.39206594228744507 -0.12889957427978516,
        3Dlights=CAD,
]{}{}{Fig5_animation.u3d}
\caption{Shaded 3D view of the {\sc SHAPE} model Homunculus with the view from Earth at the top (North is up, East is left) and a `flipped' view from the opposite direction at the bottom (North is down, East is still left). Click image for a 3D interactive view (Adobe Reader$^{\circledR}$ only).}
\label{3dview.fig}
\end{figure}


The identification of symmetric small-scale structures in the Homunculus is expected to provide clues to the ejection mechanism in relation to the existence of a highly eccentric central binary with interacting winds. We thus attempt to model small-scale features in the observed H$_2$ P-V diagrams in some detail. The features that we find with blue and red counterparts are identified in Fig.~\ref{3dview.fig} and their approximate geometric properties are summarized in Table~\ref{model.tab}.

The small-scale features have been modeled interactively by visual superposition and comparison in {\sc SHAPE}. While in principle possible \citep{N10}, automated reconstruction of high-contrast features such as those found in our observations will not work appropriately. The procedure is also limited to radial outflows. Due to the large amount of data and parameters to fit, a formal least-squares type fitting procedure is impractical. {\sc SHAPE} can do this for final adjustments of inclination angles and similar parameters. But the result is usually not significantly better than an interactive visual approach. Tests with similar features placed at random positions on an artificial objects similar to the Homunculus and then recovering the position and structure revealed a precision of azimuthal position of 5--10$^\circ$.

High contrast display of some of the P-V diagrams reveals strong deviations from a smooth lobe structure near the equatorial region (the features labeled as the Blue and Red Protrusions in Fig.~\ref{model_pv1.fig}). Our modeling shows that these are low-brightness protrusions from the main lobes located at highly symmetric positions with respect to the orientation of the binary orbit (left panel of Fig.~\ref{dents.fig}). While the detailed structure of the Blue and Red Protrusions is different from each other, they straddle the direction of binary apastron by angles of approximately 55$^\circ$ on either side, as projected onto the orbital plane. The Blue and Red Protrusions appear to be on the approaching and receding sides of orbital periastron, respectively (see e.g. fig.~12 of \citealt{madura12}). The modeling shows that the protrusions are not part of any equatorial outflow, they are clearly outside of this plane and part of the lobes (right panel of Fig.~\ref{dents.fig} and Fig.~\ref{3dview.fig}). We find that the Blue Protrusion is approximately 10$^\circ$ out of the equatorial plane, while for the Red Protrusion a specific direction cannot be determined accurately as it is spread over a range of latitudes from $\sim 10^\circ$ to 30$^\circ$.

In the polar regions there are two clearly symmetric features, in addition to smaller ``irregularities" of which the symmetry could not securely be determined. The symmetric features are the ``polar holes" and what we call ``polar trenches" (Fig.~\ref{3dview.fig}). The polar holes near the centres of the lobes are well known from spectroscopic observations (e.g. \citealt{SGK98,T08}). However, for the first time, we identify the trenches on the blue and red lobes as such. In optical observations the blue trench is easily visible as a continuous curved dark line crossing the polar region of the blue lobe. The red counterpart can only be seen in spectroscopic observations that reveal the far side of the red lobe.

The modeling reveals that the polar holes may not be completely void of dust or H$_2$. Instead of actual holes, these features may be indentations. Since the wall on the far side of the blue hole is at a large angle to the line of sight, the optical depth is lower than in other regions of the Homunculus, allowing emission from the inside to escape more easily. Unfortunately, determining whether or not these features are actual holes is difficult and depends on detailed knowledge of the exact spatial diameter of the holes, the thickness of the polar lobes, and the inclination of the Homunculus nebula \citep{T08}. While we model the holes as indentations in Fig.~\ref{3dview.fig}, within the precision of the X-shooter data and reconstruction method, models that assume a hole with an inward border of finite thickness and models assuming an indentation without a hole both reproduce the observed P-V diagrams equally well. Determining whether these features are holes in the true sense rather than indentations will require more precise measurements of the hole diameters and thicknesses of the polar lobes. Detailed investigation of the structure of the Homunculus in \FeII would also help clarify the nature of the polar holes.

The newly-identified trenches in the polar regions of the lobes are particularly striking features. They are \emph{point-symmetric} to the central stars. The modeling indicates they are not quite half-circles, centered on the polar holes and encompassing an angle of $\sim 120^\circ$ to 140$^\circ$, similar to the angular distance between the Blue and Red Protrusions discussed above. We cannot exclude the possibility that the trenches are full circles centered on the holes. The inclined nature of the Homunculus and our line of sight into the trench may produce the appearance of a partial trench. Even if this is the case, the modeling and 3D kinematic observations show that the trenches are point symmetric, with one side of the trench clearly deeper than the other.

As was the case with the polar holes, it is difficult to tell whether the trenches are complete rings without higher-resolution data. The available data and models cannot determine whether the trenches form complete circles. Whether there is any physical significance in these unexpectedly highly symmetric features is unclear at this stage. The point-symmetric nature of the trenches implies they may be unrelated to the highly eccentric central binary and are possibly a result of the originating explosion. The red and blue trenches have the same angular extent and appear to be symmetric around the apastron direction. This may be a coincidence or it may indicate some connection between these features and the central binary. How these structures might be related to circumstellar matter existing prior to the event that led to the formation of the Homunculus, or the ejection event itself, is an intriguing question for future observational and theoretical work.

\subsection{Effects of H$_2$ and dust mixing}
\label{dust.sec}

In order to better understand the observed brightness distribution in the H$_2$ data cube and its representation as P-V diagrams, we have added a dust component to the Homunculus and computed the absorption and scattering of H$_2$ with radiation transfer along the line of sight. Before introducing inhomogeneities in the H$_2$ emissivity distribution, we first investigated what brightness changes may be due to dust absorption. Therefore we used a smooth H$_2$ emissivity and dust distribution that decrease in density with distance $r$ from the central star as $\propto (r_0/(r+r_0/5))^a$, with $a=2$. This expression has been chosen as an approximation to geometric expansion with distance assuming an initially uniform density. $r_0$ is a reference distance and $r_0/5$ is used to avoid the expression to tend to infinity for small distances.

\begin{figure}
\centering \includegraphics[width=80mm]{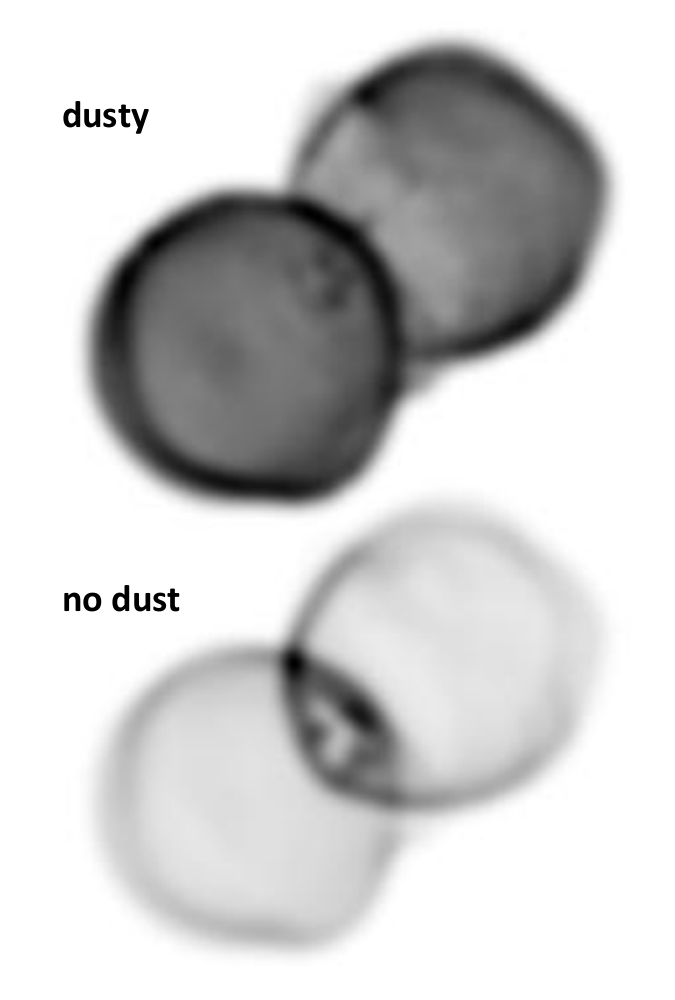}
\caption{Rendered H$_2$ model images with (top) and without (bottom) dust. The top view assumes the H$_2$ and dust are fully mixed. Since the dust-to-emission ratio decreases with distance, the image without dust is noticeably brighter in the central region.}
\label{images.fig}
\end{figure}

The dust density was set high enough to partially absorb the H$_2$ emission. We used a silicate dust radiation transfer model as described in \citet{KKS13} with the dust properties listed in Section~\ref{method.sec}. Isotropic scattering is assumed. Fig.~\ref{images.fig} shows the final rendered H$_2$ images with (top) and without (bottom) dust. In the top image, the dust and H$_2$ are assumed to be mixed.

To verify the expectation that the H$_2$ emission is mixed with the protective dust, with a thin layer of H$_2$ outside the dust shell, we considered three different cases for the relative spatial distribution of dust and molecular hydrogen. Case~1 is that of complete mixing of both components. Cases~2 and 3 are those of the dust being located mostly inside or outside of the molecular shell, respectively. For Cases~2 and 3 we generated a copy of the H$_2$ mesh, assigned a dust species to it and scaled it in such a way that the dust is either inside or outside the H$_2$ shell. The density of the dust was adjusted such that for each case the best match was found.

\begin{figure}
\centering \includegraphics[width=80mm]{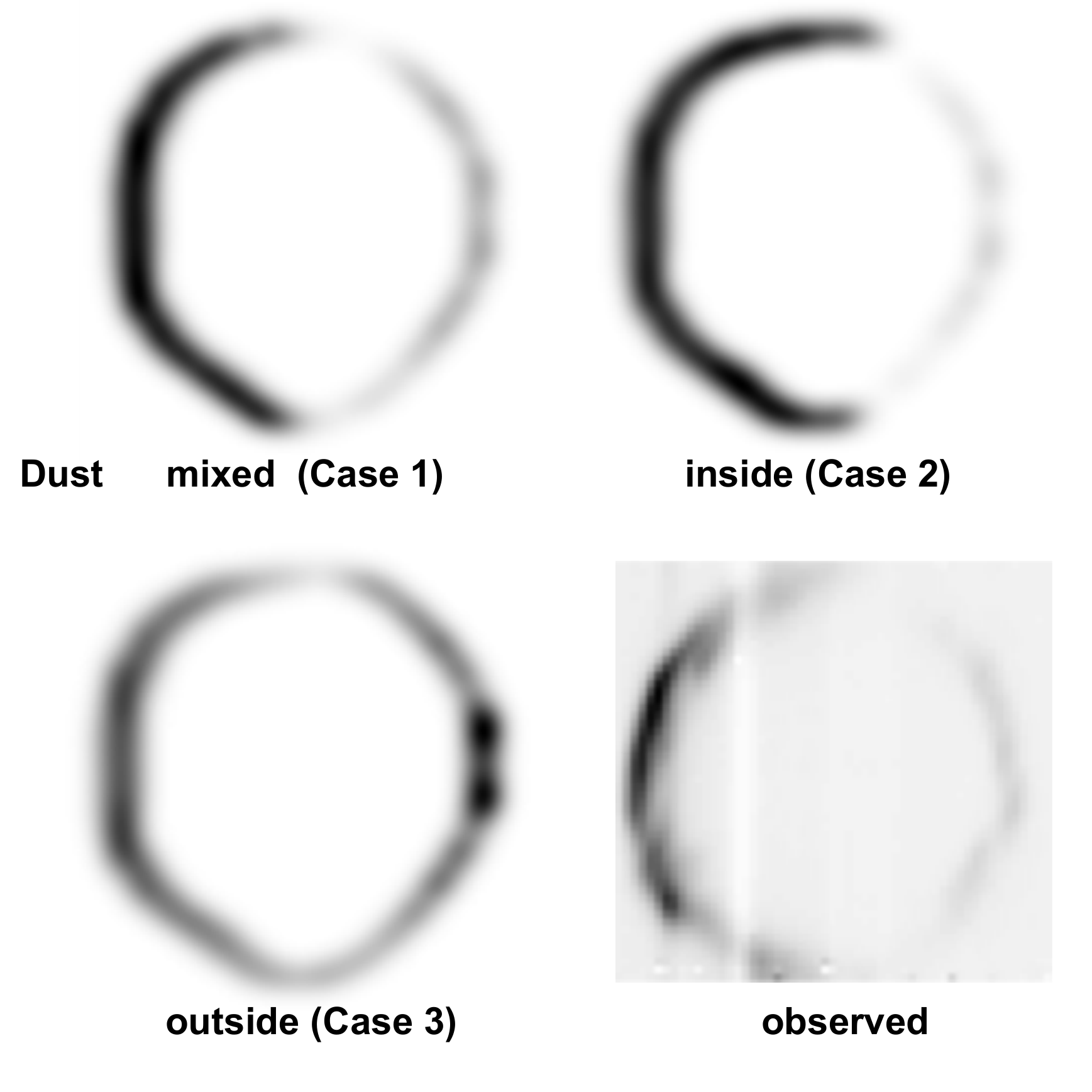}
\caption{Example {\sc SHAPE} model P-V diagrams illustrating the three different cases investigated for the relative spatial distribution of dust and H$_2$ in the Homunculus lobes. The chosen P-V diagram corresponds to slit position 30 in Fig.~\ref{model_pv1.fig}, which is for a slice through the middle of the blue lobe (Fig.~\ref{fig:refimage}). \emph{Top left}: Dust is located primarily interior to the H$_2$ shell. \emph{Top right}: Dust and H$_2$ are completely mixed. \emph{Bottom left}: Dust is located primarily exterior to the H$_2$ shell. \emph{Bottom right}: Observed P-V diagram.}
\label{dust.fig}
\end{figure}

A comparison of the three cases (Fig.~\ref{dust.fig}) shows that the P-V diagrams for Cases~1 and 3 deviate considerably from the observations, while Case~2 is able to reproduce the most salient brightness variations. As expected and observed in the P-V diagrams, the brightness of the H$_2$ emission on the backside of the P-Vs is diminished (see also Fig.~\ref{model_pv1.fig}). Some of the gaps in the P-V diagrams of the backside are reproduced and, in the model, are caused by increased dust absorption due to higher column densities. This is an effect of the local geometry, rather than intrinsic density variations of H$_2$. Here the viewing direction is roughly tangential to the surface yielding strong absorption of the far side. Most P-V images therefore show a reduced brightness near the top and bottom edges of the velocity ellipses.

When the dust is located outside of the molecular shell, both sides of the H$_2$ Homunculus receive the same level of absorption from only the front side of the dust shell, and there is not much difference in brightness between the fronts and backs of the lobes (bottom left panel of Fig.~\ref{dust.fig}). When the dust is located interior to the molecular shell, only the backside of the H$_2$ shell is absorbed, by both the front and back of the dust shell. This produces the strongest contrast between the fronts and backs of the lobes in the P-V diagrams (top left panel of Fig.~\ref{dust.fig}). When the dust and H$_2$ are fully mixed, the front is partially absorbed by only the front dust, while the back side is fully absorbed by the front dust and partially by the back dust. In reality, the situation is likely between one where the dust is located interior to the H$_2$ shell and fully mixed with it.

Strong dust absorption leads to strong variations in the P-V brightness, which, on the scale of these observations, have been reproduced by the model without intrinsic small-scale variations of the emissivity. The P-V diagrams still have many additional brightness variations though, due to intrinsic local emission and absorption variations which have not been modeled.

\section{Discussion}
\label{disc.sec}

The improved 3D model of the Homunculus presented above provides new insights on the detailed smaller-scale structure of the bipolar lobes. Two of the most significant new results of this work are the identification of the protrusions located near the equatorial plane in both lobes and the spatially-extended polar trenches. Interestingly, both features are symmetric about the apastron direction of the binary orbit, with the angular distance between the Blue and Red Protrusions ($\approx110^\circ$) being very similar to the angular extent of each polar trench ($\approx 130^\circ$). We point out that these angles are nearly identical to the total opening angle in the orbital plane of the current wind-wind collision cavity, which is $\sim 110^{\circ}$ when using a secondary/primary wind momentum ratio $\eta \approx 0.12$ (see \citealt{madura13}).

The similarity in value between the colliding-winds-cavity opening angle and the angular separation/extent of the newly discovered small-scale features (i.e. the blue and red protrusions and polar trenches) may be a numerical coincidence.  However, a key point to keep in mind regarding the Eta Carinae binary is the system's extremely high eccentricity ($\sim 0.9$), which causes the companion star to spend most of the 5.54-year orbit on the apastron side of the system. This creates a large-scale ($\gtrsim 1600$~au) time-averaged wind cavity on the apastron side of the system whose total opening angle is again $\sim 110^{\circ}$ at nearly all phases of the orbit, except for the short time near periastron passage when the companion passes behind the primary (see e.g. figures~8 and B5 of Madura et al. 2013). Interestingly, for the binary orientation derived by Madura et al. 2012, this time-averaged cavity structure lies near the Homunculus equatorial plane and along those directions on the sky where the red and blue protrusions are observed.  The similarity of the time-averaged wind-wind-cavity opening angle and the angular extent of the polar trenches is more likely a coincidence given their apparent point-symmetric nature.

When projected onto the orbital plane, the Red and Blue Protrusions appear to be located on the apastron side of the orbit. While the polar holes (Fig.~\ref{3dview.fig}) are not a new discovery, we find that both appear to be symmetrically offset from the polar axis of the large-scale Homunculus by $\sim 8^{\circ}$ and aligned with the projected apastron direction of the orbit. However, we note that the orbital axis of the binary is also slightly offset from the polar axis of the Homunculus and nearly aligned with the projected (on the sky) direction of apastron (\citep{madura12}). Therefore, the holes may be more directly related to the binary orbital axis than the projected direction of apastron, especially since the orbital axis and polar holes are closely aligned in 3D space. Alternatively, the polar holes may be aligned with the rotation axis of the central star, assuming that the stellar-rotation and binary orbital axes are also closely aligned.

With the exception of the smaller polar ``irregularities" (Fig.~\ref{3dview.fig}), which are likely a result of various instabilities in the thin Homunculus shell (see e.g. \citealt{smith13}) and/or interaction of the expanding Homunculus with pre-existing material, all small-scale features mentioned appear to be aligned relative to the apastron direction of the central binary (although in the case of the point-symmetric polar trenches this may be just a coincidence). We therefore conclude that the interaction between the outflows and/or radiation from the central binary stars and their orientation in space has had, and possibly still has, a strong influence on the structure of the Homunculus nebula. Such interactions might help explain the non-homologous expansion and small poloidal velocity component required to properly model the blue lobe (Section~\ref{velocity.sec}).

Of particular relevance is the fast, lower-density wind from the companion star, which at $v_{\infty} \approx 3000$~km s$^{-1}$, is $\sim 5$ to 40 times faster than the expanding Homunculus, depending on the latitude. In the $\sim 170$~years since the Great Eruption, the fast secondary wind has had enough time to catch up to and interact with portions of the Homunculus expanding in our direction, most especially at low latitudes near the equator where the expansion of the Homunculus is slowest \citep{smith06}. Portions of the Homunculus facing away from us are mainly unaffected by the companion star's wind since the much denser wind of the LBV primary is in the way and the colliding wind cavity opens in our direction. We speculate that some of the observed symmetric small-scale features in the Homunculus (more specifically the Red and Blue Protrusions) may be a result of interactions with the fast secondary wind. Detailed 3D numerical-hydrodynamical simulations beyond the scope of this paper are necessary to test this hypothesis, but such an interaction may help explain another peculiar feature of the Homunculus, the thin equatorial skirt.

Just as \citet{smith06} noted, we find no detectable NIR H$_{2}$ emission from the thin equatorial skirt that is prominent in optical images of the Homunculus, implying that the skirt is of much lower density than the walls of the bipolar lobes and contains significantly less mass, $\lesssim 0.5$~\msun\ \citep{smith06}. The extremely fast and low-density secondary wind has likely collided with the near-equatorial regions of the expanding Homunculus. The large difference in velocity between the secondary wind and Homunculus should result in various instabilities at the collision interface (e.g. Rayleigh-Taylor, Kelvin-Helmholtz, non-linear thin shell). This, combined with the lack of any dense material exterior to the Homunculus walls, should disrupt the thin Homunculus shell near the equatorial region, allowing the wind of the secondary to eventually plough through it.

The above-described situation is analogous to that observed in 3D smoothed particle hydrodynamics (SPH) simulations of Eta~Carinae (see e.g. fig.~8 of \citealt{madura13}). These simulations show that while the companion and its wind are at periastron, a dense, thin shell of primary wind flows toward the observer. When the binary companion returns to the apastron side, its fast wind collides with this dense shell of primary wind. Eventually, within half of an orbital cycle, instabilities in the shell cause it to break apart, allowing the fast secondary wind to plough through it. This results in a `spray' of blobs of primary wind moving in a sea of fast, low-density secondary wind, concentrated in the orbital plane. The skirt of the Homunculus may likewise be the remains of the fast secondary wind ploughing through the thin dense shell of the Homunculus, forming a spray of material concentrated in the equatorial region.

If the secondary's wind can escape from the Homunculus's equatorial regions, then so should some of the secondary's ionizing radiation. However, the secondary's radiation would not escape uniformly in all directions, but would escape preferentially in directions of lowest density, with some regions/directions shielded from the hardest UV flux by the remaining denser blobs of material that previously composed the walls of the Homunculus. The combination of ionizing secondary radiation and low-density secondary wind in the skirt could explain why no NIR H$_{2}$ emission is observed there. Moreover, because this process occurs only on the apastron side of the binary, the equatorial skirt would appear asymmetric and localized to the side of the Homunculus facing us, just as observed in optical images.

We add that this interpretation for the formation of the equatorial skirt could help resolve another of its unexplained peculiarities, namely, the presence of younger ejecta from the 1890 eruption that appears to coexist with older ejecta in the skirt \citep{SG98,davidson01,smith04,smithtownsend07}. In order to continue ploughing through the outer, more massive Homunculus, the fast secondary wind would need to penetrate the inner Little Homunculus, which is thought to have formed during the 1890 eruption \citep{ishibashi03,smith04}. Over time, the fast secondary wind and remnants of the equatorial regions of the Little Homunculus shell would be driven outward. Eventually, the blobs of older skirt material and newer material from the Little Homunculus would mix with the secondary wind. We therefore suggest that the equatorial skirt, and possibly the Red and Blue Protrusions, may be a result of the interaction of the fast secondary wind with the Homunculus and Little Homunculus post-eruption.

Of course another possibility is that the symmetric small-scale features in the Homunculus are intrinsic to the Great Eruption. The strongest evidence for this possibility comes from the point-symmetric nature of the polar trenches and polar holes. Even if this is the case, the relation of these symmetric features to the apastron direction of the orbit and/or the binary orbital axis strongly implies that binarity had some role in helping trigger the eruption and/or shape the nebula. More evidence for a link between binarity and the Great Eruption comes from the correlation of periastron passage with the brightening peaks observed in Eta~Carinae's light curve in the decades leading up to the eruption (see fig.~2 of \citealt{damineli96} and also \citealt{smithfrew11}). A stellar collision scenario such as that envisioned by \citet{smith11} or a binary interaction like that in \citet{kashi10} may have occurred during the eruption, although the details are far from certain.

Another pair of peculiarities worth discussing are the polar holes of the Homunculus. Are these true holes or just indentations in the lobes? Were the holes created during the original eruption event, or significantly later afterward? What physical process formed them? The holes may be the result of instabilities in the expanding bipolar structure, similar to those that likely created the smaller irregularities seen in the NW red lobe. However, one would have to explain why such instabilities produced a pair of point-symmetric holes that are offset from the polar symmetry axis of the nebula. Instead, perhaps some sort of polar jet closely aligned with either the central star's rotation axis or the binary's orbital axis was present and produced the holes. This jet may have been present during the initial eruption or created much later. Unfortunately, we do not have enough information to constrain the timing of any hypothetical jets, nor can we currently explain the potential mechanisms behind their formation (see e.g. \citealt{soker03}). Detailed examination of the Homunculus in \FeII emission and of the interior Little Homunculus, searching for the presence of similar polar holes, would help greatly in understanding their origin.

One final note regarding the polar holes. It would be interesting to investigate what the existence of these features imply for theories for the formation of the Homunculus that rely on mass-loss from a rapidly-rotating star \citep{owocki97,owocki98,dwarkadas02,owocki03,smith06}. In such models, the mass-loss, and hence density, is expected to be highest over the poles. Yet, the mass fraction as a function of latitude in the lobes clearly shows a peak between $50^{\circ}$ and $60^{\circ}$ (see fig.~5b of \citealt{smith06}), with a much smaller mass fraction at the poles. Combined with the difficulties of a steady radiation-driven wind powering such extreme mass-loss \citep{owocki04, smithtownsend07}, this may indicate that steady mass-loss from a rapidly rotating single star is not the correct explanation.

The newly-discovered point-symmetric polar trenches and near-equatorial Red and Blue Protrusions, as well as the polar holes, pose some interesting complications for models for the Homunculus that are based on an explosion from a rapidly-rotating single star. Prior to the observations and 3D model in this paper, the model by \citet{smithtownsend07} seemed a potentially viable explanation for the shape of the Homunculus. However, that model produces an axisymmetric nebula and skirt that do not match the more detailed structures apparent in the new observations presented above. We emphasize that this does not rule out the possibility that the central primary star was near critical rotation for mass loss or that some sort of explosion occurred which helped power the Great Eruption. Instead, it more strongly implies that single-star models for the formation and shape of the Homunculus are not correct, and that binarity and interacting stellar winds played an important role in shaping the detailed 3D morphology of the Homunculus. It would be interesting to see if the inclusion of binarity and colliding winds in a model similar to that of \citet{smithtownsend07} can produce the more detailed 3D structure of the Homunculus.

As one can clearly see, the Homunculus is an extremely complex structure whose formation and shape will likely only be adequately explained with the assistance of 3D numerical-hydrodynamical simulations. Numerous possible scenarios exist for triggering the eruption and shaping the large- and small-scale features observed. Future models should strongly consider the effects of binarity, collisions, and the spatially-extended colliding stellar winds. Stellar mergers should also be explored. The pre-eruption environment immediately surrounding the central stars is also expected to be of some importance in influencing the resultant asymmetric shape and various possible instabilities, and should be included if possible.

\section{Conclusions}
\label{conclusions.sec}

In this paper we presented new kinematic molecular hydrogen data from the ESO VLT/X-shooter instrument that fully map the Homunculus nebula around Eta~Carinae. Based on these observations we constructed the first full 3D model of the shape of the bipolar nebula that includes small-scale structures. As in \citet{smith06}, we find no detectable NIR H$_{2}$ emission from the thin equatorial skirt that is prominent in optical images of the Homunculus. By introducing a non-radial correction to the velocity field, we find a unique solution for the large-scale structure of the bipolar Homunculus, unifying earlier attempts that yielded different solutions for models based on imaging and kinematic data.

We further extended the model to incorporate observed small-scale deviations from axisymmetry. The geometric positioning of these features is summarized in Table~\ref{model.tab}. Based on our observations, we accurately place the positions of the well-known polar holes, finding that they appear to be symmetrically offset from the axis of the large-scale Homunculus by approximately $8^\circ$. In both cases this deviation is closely aligned with the apastron direction of the central binary orbit and the 3D orientation of the binary's orbital axis.

Similarly, we find that the newly identified polar trenches (Fig.~\ref{3dview.fig}) are point-symmetric with respect to the central stars and may also be related to the apastron direction of the binary orbit. The most intriguing new features are the Blue and Red Protrusions in the lobes (Figs.~\ref{model_pv1.fig} - \ref{3dview.fig}). They are not exactly in the equatorial plane, but of order $10^\circ$ in latitude out of the plane. In longitude they are both about $55^\circ$ away from the direction of apastron, in opposite directions. It might be noteworthy that the angular distance between the Blue and Red Protrusions ($\approx 110^\circ$) is similar to the angular extent of each of the polar trenches ($\approx 130^\circ$), and nearly identical to the estimated total opening angle of the current-day inner wind-wind collision cavity ($\approx 110^\circ$).

The smaller ``irregular" features in Fig.~\ref{3dview.fig} have no apparent symmetry or relation to the orientation of the binary orbit. They might be attributable to instabilities in the expanding Homunculus. All other major small-scale feature appear to be aligned relative to the apastron direction of the central stars, although in the case of the point-symmetric polar trenches this may be a coincidence. We therefore conclude that the interaction between the outflows and/or radiation from the central binary stars and their orientation in space has had, and possibly still has, a strong influence on the structure of the Homunculus nebula. Interaction of the fast wind from the secondary star with the near-equatorial regions of the Homunculus and Little Homunculus may help explain the protrusions and thin equatorial skirt. The possibility that the current stellar winds are influencing the Homunculus's structure, plus the requirement of a non-homologous expansion to properly model the SE blue lobe, imply that the standard assumption of a homologous expansion may no longer hold.

Our new observations and 3D model pose serious complications for single-star theoretical models of the Homunculus's formation. Prior to this work, there was no strong reason to consider a binary model for the formation or shape of the Homunculus. Detailed 3D numerical-hydrodynamical modeling of the Homunculus's formation that includes the central binary and its interacting winds are now required to determine if such interactions can simultaneously reproduce the newly-identified small-scale features, overall large-scale bipolar shape, the polar holes, and thin equatorial skirt.

\section*{Acknowledgments}
W.~S. acknowledges financial support through grant UNAM-PAPIIT IN101014. M.~T. is supported by CNPq/MCT-Brazil through grant 201978/2012-1. T.~I.~M. is supported by an appointment to the NASA Postdoctoral Program at the Goddard Space Flight Center, administered by Oak Ridge Associated Universities through a contract with NASA.  J.~H.~G. is supported by an Ambizione fellowship of the Swiss National Science Foundation. A.~D. acknowledge FAPESP for continuous financial support. The authors thank D.~Clark for useful discussions.

\bsp

\label{lastpage}

\end{document}